\documentclass[tighten,twocolumn]{aastex701}
\usepackage[]{graphicx}
\usepackage{amsmath}
\usepackage[normalem]{ulem}

\shortauthors{Kerr}
\shorttitle{On Gamma-Ray PTAs}

\begin{document}

\title{Future Space-based Gamma-ray Pulsar Timing Arrays}

\author[0000-0002-0893-4073,gname=Matthew,sname=Kerr]{M.~Kerr}
\affiliation{Space Science Division, Naval Research Laboratory, Washington, DC 20375--5352, USA}
\email{matthew.kerr@gmail.com}
\correspondingauthor{M.~Kerr}

\author[0000-0002-9249-0515, gname=Zorawar,sname=Wadiasingh]{Z.~Wadiasingh}
\affiliation{Department of Astronomy, University of Maryland, College Park, MD, 20742, USA}
\affiliation{Astrophysics Science Division, NASA Goddard Space Flight Center, Greenbelt, MD 20771, USA}
\affiliation{Center for Research and Exploration in Space Science and Technology, NASA/GSFC, Greenbelt, MD 20771, USA}
\email{zwadiasingh@gmail.com}

\author[0000-0003-1521-7950,gname=Adrien,sname=Laviron]{A.~Laviron}
\affiliation{Astrophysics Science Division, NASA Goddard Space Flight Center, Greenbelt, MD 20771, USA}
\email{adrien.laviron@nasa.gov}

\author[0000-0003-1080-5286,gname=Constantinos,sname=Kalapotharakos]{C.~Kalapotharakos}
\affiliation{Astrophysics Science Division, NASA Goddard Space Flight Center, Greenbelt, MD 20771, USA}
\email{konstantinos.kalapotharakos@nasa.gov}

\author[0000-0002-6039-692X,gname=H.~Thankful,sname=Cromartie]{H.~T.~Cromartie}
\affiliation{National Research Council Research Associate, National Academy of Sciences, Washington, DC 20001, USA resident at Naval Research Laboratory, Washington, DC 20375, USA}
\email{thankful.cromartie@nanograv.org}

\author[0000-0001-7587-5483,gname=Tyler,sname=Cohen]{T.~Cohen}
\affiliation{Department of Physics, New Mexico Institute of Mining and Technology, 801 Leroy Place, Socorro, NM 87801, USA}
\email{tyler.cohen@nanograv.org}

\begin{abstract}
Radio pulsar timing array (PTA) experiments using millisecond pulsars (MSPs) are beginning to detect nHz gravitational waves (GWs).  MSPs are bright strong GeV $\gamma$-ray emitters, and all-sky monitoring of about 100 MSPs with the Fermi Large Area Telescope (LAT) has enabled a $\gamma$-ray Pulsar Timing Array.  The GPTA provides a complementary view of nHz GWs because its MSP sample is different, and because the $\gamma$-ray data are immune to plasma propagation effects, have minimal data gaps, and rely on homogeneous instrumentation.  To assess GPTA performance for future $\gamma$-ray observatories, we simulated the population of Galactic MSPs and developed a high-fidelity method to predict their $\gamma$-ray spectra.  This combination reproduces the properties of the LAT MSP sample, validating it for future population studies.  We determined the expected signal from the simulated $\gamma$-ray MSPs for instrument concepts with a wide range of capabilities.  We found that the optimal GPTA energy range begins at 0.1--0.3\,GeV (depending on angular resolution) and extends to 5\,GeV. We also examined concepts operating in the Compton (MeV) regime. With the caveat that the MSP spectra models are extrapolated beyond observational constraints, we found low signal-to-background ratios, yielding few MSP detections.
GeV-band concepts would detect 10$^3$ to 10$^4$ MSPs and achieve GW sensitivity on par with and surpassing the current generation of radio PTAs, reaching the GW self-noise regime.     When considering two possible scenarios for the formation of MSPs in the Galactic bulge---the collective signal from which is a potential source of an excess GeV signal observed towards the Galactic center---we find that most of the concepts can both detect this bulge population and distinguish the production channel.  In summary, the high discovery potential, strong GW performance, and tremendous synergy with radio PTAs all argue for the pursuit of next-generation $\gamma$-ray pulsar timing.
\end{abstract}


\section{Introduction}




The relative tick frequency of separated clocks varies according to the intervening spacetime metric, so measurements of such frequency variations can be used to infer the presence of gravitational waves (GWs).
When the clock-observer separation is large compared to the
GW wavelength, the frequency shift depends only on the
GW amplitude at the two positions \citep{Estabrook75}.
This is the case for nHz GWs when the clocks are realized with pulsars, which have a typical distance of 1\,kpc.  These amplitudes are then known as the pulsar and earth terms.  In
general, the GW phase at the pulsar is not known, but the GW phase at the earth is common
for all pulsar observations.  Consequently, monitoring many
pulsars and correlating the resulting data can reveal the presence and
properties of the GW through the earth term \citep{1979ApJ...234.1100D}.
The angular correlations
of the earth term between the pulsars is uniquely determined by the
quadrupolar antenna pattern \citep{Hellings83}.

Pulsar timing arrays (PTAs) are observational campaigns designed to
realize an ensemble of clocks by making high-precision ($\sim$$\mu$s) measurements of
the times of arrival (TOAs) of pulses from millisecond
pulsars\footnote{On these timescales, the timing precision achievable with MSPs is several
orders of magnitude better than can be achieved with young/unrecycled
pulsars.} \citep[MSPs,][]{1982Natur.300..615B,Alpar82} distributed across the sky.  Timing models account
for the unique properties of each pulsar (spin frequency, position,
etc.) and enable the prediction of TOAs relative to the terrestrial timescale.  The
differences between predicted and observed TOAs are a realization of near-ideal, separated clocks, and are the starting point for the analysis of
GWs.

The strain amplitude $h$ of GW from a near-periodic source with frequency $f$ is often given as
a dimensionless characteristic strain $h_c \propto fh$, and power produced in the detector is $\propto h^2 \propto h_c/f^2$ \citep{Moore15}.  Meanwhile the noise power in GW
detectors is set by the error in clock tick measurements, or phase noise.  If the noise power is independent of frequency (white), then the minimum detectable $h_c\propto f^2$, i.e. in the absence of other noise,
detectors are more sensitive to lower-frequency GW.  Terrestrial GW detectors like LIGO operate in the audio
band and are limited at low frequencies by environmental perturbations of the test masses, so that instead the $h_c(f)$ sensitivity curve has a characteristic bucket-like shape.  LISA will operate in the lower 1\,mHz band, but its sensitivity has a
similar shape, with the low-frequency limit set by Brownian noise and environmental effects in the gravitational reference sensor surrounding the test mass \citep{2024arXiv240207571C}.

Although PTAs are not interferometers, they too have a bucket-shaped
sensitivity \citep[e.g.][]{Agazie23_NG15_detector_characterization}.  The phase noise is set by the measurement precision on
TOAs.  The low-frequency turnup is mostly influenced not by noise but by the absorption of the GW signal into timing
model parameters because the pulsar period $P$ and spindown rate $\dot{P}$ are unknown a priori.  This means that the optimal sensitivity for PTAs is
achieved at roughly half the length of the data span,
$T_{\mathrm{obs}}/2$.  For PTAs, $T_{\mathrm{obs}}$ is of order 10
years, so PTAs are most sensitive to nHz GWs.

Although there are more exotic GW sources proposed in this band \citep[e.g.,][]{2023ApJ...951L..11A}, including those associated with the early universe~\citep{1988PhRvD..37.2078A,1997rggr.conf..373A}, 
the loudest PTA GW sources are expected to be merging binary
supermassive black holes (SMBHBs).  In the nHz band, sources evolve
slowly and are nearly monochromatic on 10-year timescales.  Meanwhile, the
angular resolution of PTAs is limited both formally \citep{Boyle12}
and practically\footnote{The universe has only provided a fixed number of
MSPs and they have extremely heterogeneous timing precision.}.
Effectively this means that the superposed signals from merging SMBHBs
throughout the universe are confused, with only atypically loud sources possibly forming hotspots.  Consequently, the initial science driver for PTAs is the detection and characterization of the combined
background hum, the stochastic GW background (GWB).  Continued studies with more sensitive detectors may reveal individual sources\footnote{If the universe cooperates, it may be possible even to measure pulsar terms and phase up PTAs \citep{2022MNRAS.517.1242M}.  This requires two or more nearby SMBHBs sources near the end of their inspiral phase, with a GW frequency that evolves over $T_{\mathrm{obs}}$.  Determining the GW phase allows pulsar distance measurements to a fraction of the GW wavelength, $0.1-1$ parsec, for sufficiently bright MSPs. Once the pulsar distances are known, it increases the sensitivity of the PTA to all other nHz GW signals!}.

The simplest version of the GWB for a smooth, cosmological distribution
of inspiraling circular binaries predicts a gravitational
wave spectrum $h_c(f) = \mathcal{A}_{\mathrm{gwb}}
(f/\mathrm{yr}^{-1})^{-\alpha}$, with $\alpha=2/3$ for the canonical
SMBHB merger scenario \citep{2001astro.ph..8028P}.  The normalization depends on the chosen cosmology, and the number
density and mass spectrum of SMBHB members, while deviations from this simple
prediction are expected from the astrophysics of SMBHB mergers and population
evolution \citep[e.g.][]{Burke-Spolaor19,2025arXiv250500797K}.

Such a stochastic GW spectrum will induce a stochastic
noise process in pulsar timing residuals with a steep, red power spectrum given
by $P(f) =
\mathcal{A}_{\mathrm{gwb}}^2/(12\pi^2)(f/\mathrm{yr}^{-1})^{-\Gamma}
\mathrm{yr}^{3}$, with $\Gamma={13/3}$ for $\alpha=2/3$.  Although
other noise processes, including intrinsic ``spin noise'' are also present in MSPs \citep{Shannon10,Reardon23}, the
stochastic GWB is a noise floor that is common to every pulsar.  Thus,
PTAs seek to monitor many MSPs with both good timing precision and low intrinsic noise in order to detect this common signature.

PTAs have historically made use of large radio telescopes, often
single dishes or arrays thereof, with relatively narrow
fields-of-view.  Observing bandwidths are as large as possible in order to reduce measurement noise and to characterize delay from cold plasma dispersion in the intervening ionized interstellar medium (IISM), $\delta \mathrm{TOA}\propto \mathrm{DM}\,\nu^{-2}$, where the dispersion measure (DM) is the electron column density.  DM varies with time due to mutual motion of the earth and pulsar, and the induced TOA variations dwarf other noise and signals, so frequent wideband monitoring is required to measure DM \citep{Keith13}. The resulting data sets are heterogeneous and complex, and IISM models are ad hoc and may fall short of reality.  There is not yet a widely accepted optimal method for mitigating IISM propagation effects, of which DM variations are the simplest \citep{Iraci24}.

\defcitealias{Smith23}{3PC}
One alternative approach is to observe in wavelengths that are unaffected by the IISM, and both young pulsars and MSPs emit the bulk of their radiation in the
$\gamma$-ray band.  Indeed,
the Fermi Large Area Telescope \citep[LAT, ][]{Atwood09} has detected
and now continuously monitors more than 100 MSPs \citep[3PC,
][]{Smith23}.  Although its collecting area is small compared to radio
telescopes, the LAT enjoys certain advantages in pulsar timing: it
simultaneously observes every accessible MSP in its 2-steradian field-of-view;
the $\gamma$-ray signals from MSPs are unperturbed by the IISM; data analysis is more straightforward with far fewer nuisance model parameters; and the LAT data set approaches two decades in length, with
negligible interruption or change in experimental setup.

In the last few years, radio PTAs have inched towards a detection of
the GWB.  Initial reports focused on the prevalence of red noise
that seemed to have a common amplitude and spectral shape across the
pulsar population and was consistent with the predictions for a GWB
from SMBH mergers.
Subsequent reports began to reveal the expected angular correlations in TOAs
between pulsars: see \citet{2024ResPh..6107719V} for a comprehensive overview.
Consequently, it is reasonable to conclude that the stochastic GWB is
detectable and that is has an amplitude of about
$\mathcal{A}_{\mathrm{gwb}}\sim2\times10^{-15}$, or $\mathcal{A}_{15}=2$. 

In parallel, \citet{Ajello22} pioneered the use of $\gamma$ rays for pulsar
timing arrays, in particular developing the necessary methodology to detect a GWB
using sparse photon data instead of TOAs\footnote{It can take the LAT months or even years to collect enough photons to make a precise phase measurement on a given pulsar.}.  Using these techniques and a sample of 35 $\gamma$-ray MSPs, \citet{Ajello22} reported an
upper limit on the GWB of $\mathcal{A}_{15}\leq10$ and projected that with
accumulated data, the $\gamma$-ray Pulsar Timing Array (GPTA) could detect a
GWB with $\mathcal{A}_{15}\sim2$ by about 2030.

The GPTA, currently based only on LAT data, can play a key role in PTA science in the
coming years.  Its best-performing pulsars are different to those in
radio PTAs, and it offers the opportunity to characterize spin noise
in the absence of IISM effects.  While its accumulating data set makes
it an ever more successful PTA, it is tempting to entertain what could be
accomplished with an even more capable instrument, perhaps even one (unlike the
LAT) designed from the outset for PTA science goals.  That forecast
is the goal of this work.

In making our calculations, we necessarily predict
the number of future $\gamma$-ray MSP detections, and these results
indicate the potential to discover 10$^3$--10$^4$ new $\gamma$-ray
MSPs.  Many of these MSPs are likely to be radio MSPs, and current
experience shows that directing radio telescopes at MSP-like
$\gamma$-ray sources \citep{Ray12} is a powerful mechanism for discovering new MSPs
in both bands \citep[e.g.,][]{Ransom11,Kerr15,Bangale24,Kerr25}.  Thus, in addition to providing independent and complementary GW and spin noise measurements, a future $\gamma$-ray PTA would
also discover an immense number of MSPs and improve both radio and $\gamma$-ray PTA samples. 

To make our PTA performance assessments, we first generate ensembles of MSP
populations following the scheme laid out in \S\ref{sec:popsynth}.  To
these MSPs we attach a complete $\gamma$-ray spectral model with parameters that depend
on the spindown luminosity and magnetic field of each MSP, as
described in \S\ref{sec:gammamodel}.  We then develop a prescription for
computing the sensitivity of future instruments to these simulated
MSPs by scaling to the current performance of the LAT in 
\S\ref{sec:virtuallat}, and we further derive a mapping between point
source sensitivity and pulsar timing performance in \S\ref{sec:timing}.
In \S\ref{sec:ptaperf}, we combine the single-pulsar timing
performance into an assessment of PTA performance as characterized by
sensitivity to the GWB.

Our primary results begin in \S\ref{sec:gev}, where we consider
instruments that, like the LAT, operate in the $e^{\pm}$ pair production regime.  We demonstrate
that these instruments have by far the most potential for PTA
purposes.  However, we also consider Compton regime instruments in
\S\ref{sec:mev}.  Finally, we discuss our results and offer
guidance on the possible choices for implementing future
$\gamma$-ray PTA instruments in \S\ref{sec:discussion}.

\section{Millisecond Pulsar Population Synthesis}
\label{sec:popsynth}

\begin{figure}
\centering
  \includegraphics[angle=0,width=1.00\linewidth]{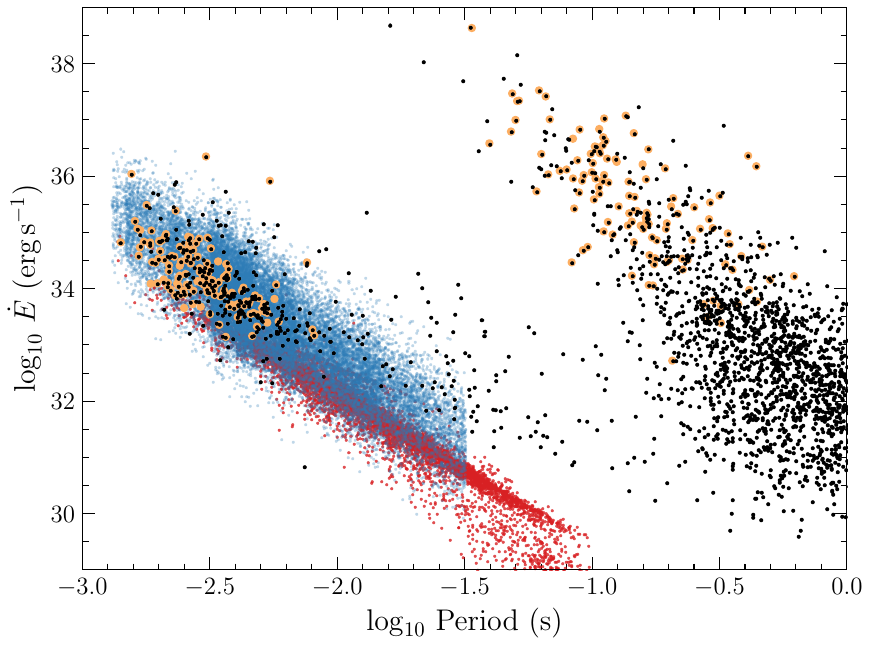}
  \caption{\label{fig:popsynth}The distribution in $P$ and $\dot{E}$
  for various populations of MSPs and young pulsars.  The black points are pulsars tabulated in
  the ATNF pulsar catalog
  \citep{Manchester05}\footnote{http://www.atnf.csiro.au/research/pulsar/psrcat}
  and the orange points are
  LAT-detected pulsars.  The blue cloud of points is one realization
  of the disk population of MSPs (S1) discussed in the text, while the red
  cloud shows the distribution of a possible bulge population of MSPs
  formed via accretion-induced collapse (S2).  The MSPs of another bulge population scenario, S3, follow the same distribution as S1. The period cutoffs of 30\,ms (S1/S3) and 100\,ms (S2) that were assumed in the synthesis are evident, but these cutoffs are irrelevant for PTAs, which prioritize rapid rotators.}
\end{figure}

The MSP population in our galaxy exists as multiple components in the disk, bulge, globular clusters and halo, tracing the complex stellar histories of the local environments. As MSPs are binary stellar evolution products, their formation depends on the stellar encounter rate, distribution of binary separations and component masses, kicks from the natal supernova, the local Galactic potential and their movement through it, and probability of retention in the galaxy \citep[e.g.,][]{1999A&A...350..928T,2007ApJ...671..713S,2008MNRAS.388..393K,Gonthier18,2020ApJS..247...48K,2022ApJ...934L...1K,Shawaiz25,2025arXiv251015661S}. Given this complex history, most MSP population studies adopt an empirical approach, calibrating parametric spatial distribution and pulsar properties with observational surveys, first attempted by \citet{1990ApJ...348..485P}.  We consider three scenarios for the Galactic population: S1 contains only disk MSPs, while S2 and S3 contain the disk MSPs of S1 and an additional bulge component.  In all analyses, we analyze 10 realizations of each population.

\subsection{The Disk MSP Population: S1}

We synthesize the population of MSPs in the Galactic disk using
\texttt{PSRPOPPy} \citep{Bates14}, which generates populations
from analytic distributions in spin period $P$,
period derivative $\dot{P}$, and position, e.g. following a gaussian
disk profile with a certain width and scale height.  The specific
choice of parameters and functions is given in \citet{Liu23}.  By construction, the spatial distribution follows the Galactic disk,
i.e., there is no explicit bulge population.  The distribution of $P$
and spindown luminosities\footnote{For consistency with the
literature, we adopt canonical neutron star scalings: a radius of
10$^6$\,cm and a moment of inertia of $10^{45}$\,g\,cm$^2$.}
$\dot{E}\equiv I (2\pi)^2 \dot{P}/P^3$\,erg\,s$^{-1}$ for a single
realization is shown in Figure \ref{fig:popsynth}.
The simulated population agrees in bulk with the observed MSP
population, though it includes an extrapolated component towards lower
$\dot{E}$ that is not (yet) reflected in MSP surveys at the current
sensitivity limits.  It is also apparent that, for a given $P$, the
range of $\dot{E}$ extends to larger values than the observed
$\gamma$-ray MSP population\footnote{However, some radio MSPs appear in this $\dot{E}$ range, presumably in globular clusters with a distance that precludes detection with the LAT.}.


We additionally considered a realization of the population with
parameters determined by \citet{Shawaiz25}, who tuned the population
parameters to agree with the number of detected LAT MSPs and
unassociated point sources that could be undetected MSPs.  However,
when coupled with the $\gamma$-ray emission prescription detailed below,
this population produced a logN-logS distribution that was too
shallow, so we did not adopt it.

\begin{figure*}
\centering
  \includegraphics[angle=0,width=0.49\linewidth]{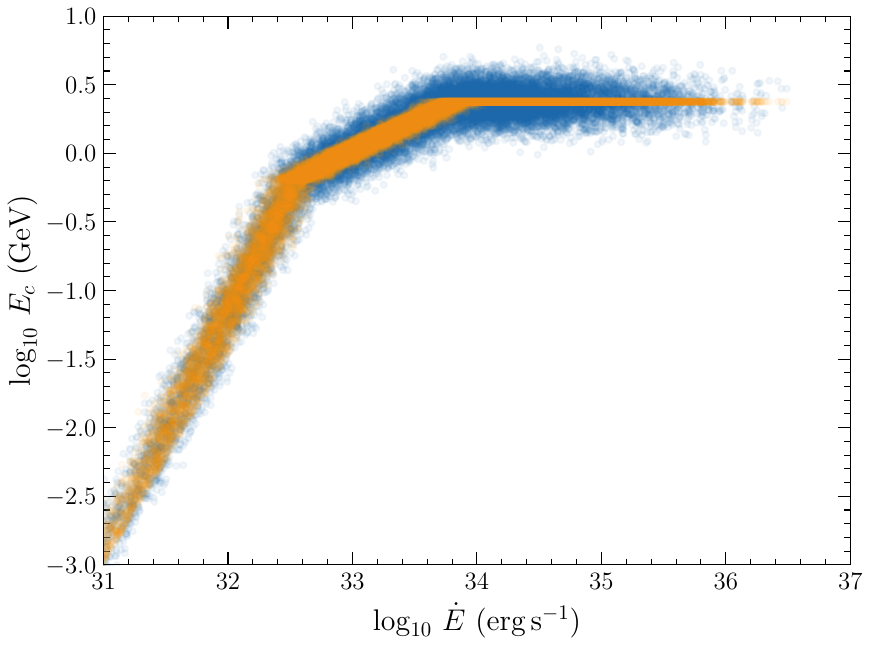}
  \includegraphics[angle=0,width=0.49\linewidth]{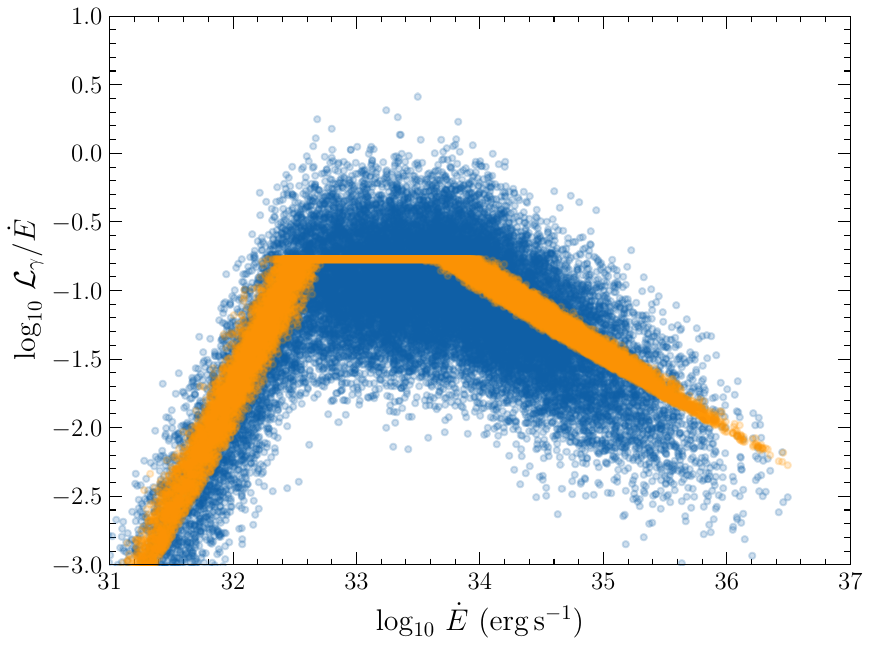}
  \caption{\label{fig:fp}The predicted cutoff energy ($E_c$, left) and
  $\gamma$-ray efficiency (right) distributions for one realization of the disk population with the
  $\gamma$-ray spectrum derived using the fundamental plane
  relation outlined in the main text.  The orange points indicate the direct application of the relation to the synthesized $P$ and $\dot{P}$, and the blue points indicate the introduction of scatter on both $E_c$ and $\dot{E}$.  The former captures the observed observational spread in $E_c$, and the latter captures beaming and other neglected physical influences on the luminosity (e.g. magnetic inclination).}
\end{figure*}

\subsection{MSP Bulge Populations}
\label{sec:bulge}

A population of MSPs that traces the
Galactic bulge might originate from several mechanisms, and the
Galactic Center Excess \citep[e.g.][]{Calore15} observed in LAT data could
plausibly originate from the summed, unresolved emission from such a
population \citep[e.g.,][]{2020JCAP...12..035P,2024PhRvD.109l3042M}.  At $\sim$8\,kpc, these pulsars are
generally too faint to be individually detected by the LAT or by radio pulsar
surveys.

A future, sensitive $\gamma$-ray instrument could detect bulge MSPs,
and so we consider two scenarios to assess the effect of such
populations on $\gamma$-ray pulsar timing arrays.  In both scenarios,
the spatial distribution is chosen to follow the GCE, specifically a
power-law distribution $N(r)\propto R^{-2.56}$ \citep{Clark25}, which is essentially the square of the best-fitting generalized Navarro-Frenk-White \citep{Navarro96} profile \citep{Calore2015}.  These scenarios are thus more concrete realizations of the ad hoc luminosity functions explored by \citet{2022JCAP...06..025D}.

\subsubsection{Accretion-induced collapse: S2}
In the first bulge scenario, we choose a population of MSPs formed by the
accretion-induced collapse (AIC) of white dwarfs \citep{Hurley10}.  Here, the high angular momentum and magnetic field originate from an assumed conservative collapse, though additional mass transfer can further spin up the MSP.  Because these MSPs would have formed during early
starburst activity over 10\,Gyr ago, the subsequent spindown yields a
substantially lower typical $\dot{E}$, at present, compared to the
disk MSPs.  We use the $P$ and $\dot{P}$ values from the binary
synthesis of \citet{Gautam22}, who explored this population in the
context of the GCE, using each pulsar 14 times in order to scale from the initial mass function and to reproduce the
luminosity of the GCE.  This yields a population of $1.2\times10{^5}$
MSPs as shown in Figure \ref{fig:popsynth}. This number is similar to that required by the analysis of \citet{2025arXiv250717804L}.

\subsubsection{A disk-like bulge: S3}
In the second scenario, we simply let the bulge MSPs follow the same
$P$ and $\dot{P}$ population as the primary disk population.
Normalization to the GCE requires about $10^4$ such pulsars
\citep{Gonthier18}, and when we use the $\gamma$-ray luminosity mechanism below, we find the precise normalization is 6,700 MSPs.  We adopt this value for S3 population.  Actual detection rates would scale linearly with the assumed value, but our main conclusions are insensitive to the precise value.

\section{Determining Gamma-ray Emission}
\label{sec:gammamodel}

The phase-averaged $\gamma$-ray luminosity, $L_{\gamma}$, of pulsars is empirically
observed (see \citetalias{Smith23}) to scale roughly as $\dot{E}^{1/2}$, and this
relation is readily explained by considering the available polar cap
voltage \citep{1996A&AS..120C..49A,2013ApJS..208...17A}.  \citet{Kalapotharakos19,Kalapotharakos22,2023ApJ...954..204K} identified a more
detailed fundamental plane-type (FP) relation that further links
$L_{\gamma}$, $\dot{E}$, the surface magnetic field
($B\equiv\frac{3c^3\,I}{8\pi^2\,R^6}P\dot{P}$), and the spectral
cutoff energy $E_c$.  Since $P$ and $\dot{P}$ are tabulated for each
simulated MSP, we can evaluate $\dot{E}$ and $B$, so using the FP relation, we can determine a properly-normalized phase-averaged
$\gamma$-ray spectrum if we can assign a reasonable $E_c$ to each MSP\footnote{The reverse also appears to be true, and distances of radio-quiet pulsars can be estimated using the FP \citep{2025arXiv250510208O}.}.


Following \citet{2025kalapotharakos}, we adopt the two-branch equatorial-current-sheet curvature prescription for the spectral cutoff energy. For high-$\dot{E}$ pulsars, the primary particle energies are limited by radiation reaction from curvature radiation, yielding $E_c = 10.2\,\mathrm{GeV}\,\eta_{R\mathrm{LC}}^{1/2}\eta_{B\mathrm{LC}}^{3/4}\eta_{\alpha}^{-7/16}\dot{E}_{36}^{7/16}B_{12}^{-1/8}R_6^{-3/8}$, where $\eta_{R\mathrm{LC}}$ is the curvature radius in units of $R_{\mathrm{LC}}=c P/(2 \pi)$ (the light cylinder), $\eta_{B\mathrm{LC}}$ the accelerating electric field scaled to $B_{\mathrm{LC}} = B/R_{\mathrm{LC}}^3$, and $\eta_{\alpha}$ the variation of spin-down power with magnetic inclination; $\dot{E}_{36}\equiv\dot{E}/10^{36}\,\mathrm{erg\,s^{-1}}$, $B_{12}\equiv B/10^{12}\,\mathrm{G}$, and $R_6=r_\star/10\,\mathrm{km}$. For lower $\dot{E}$ pulsars, below the transitional spin-down power $\dot{E}_{\mathrm{TR}}$, where the radiation-reaction Lorentz factor equals that set by the available potential drop, the maximum particle energy becomes limited by the total polar-cap potential, giving $E_c = 2.8\times10^6\,\mathrm{GeV}\,\eta_{\mathrm{pc}}^3\eta_{R\mathrm{LC}}^{-1}\eta_{\alpha}^{-7/4}\dot{E}_{36}^{7/4}B_{12}^{-1/2}R_6^{-3/2}$, with $\eta_{\mathrm{pc}}$ the fraction of the polar-cap voltage tapped by the primaries. Although the scale factors introduce mild degeneracies, the joint constraints from Fermi-LAT observations and global 3D particle-in-cell simulations confine them to narrow, physically plausible ranges. We adopt $\eta_{B\mathrm{LC}}=0.25$, $\eta_{R\mathrm{LC}}=6.6$, $\eta_{\mathrm{pc}}=0.4$, and $\eta_{\alpha}=3/2$ \citep[see][]{2025kalapotharakos}. The transition condition produces a knee in the $\gamma$-ray luminosity at $\dot{E}\simeq7.3\times10^{31}\,\mathrm{erg\,s^{-1}}\,R_6^{6/7}B_{12}^{2/7}\eta_{R\mathrm{LC}}^{8/7}\eta_{B\mathrm{LC}}^{4/7}\eta_{\mathrm{pc}}^{-16/7}\eta_{\alpha}\approx3\times10^{32}\,\mathrm{erg\,s^{-1}}$. 

For reference, $E_c$ is about 1\,MeV for MSPs with
$\dot{E}=10^{31}$\,erg\,s$^{-1}$, which is about the lowest $\dot{E}$
observed in the disk population.  However, bulge MSPs from the
accretion-induced collapsed synthesis reach
$\dot{E}=10^{30}$\,erg\,s$^{-1}$ (implied $E_c\approx 40$\,keV) and
below. Such systems may not sustain a near-force-free magnetosphere and, in any case, fall below the $\gamma$-ray visibility threshold; we therefore retain them in the population, but their contribution to detectable emission is negligible. 

At the opposite end, \citet{2025kalapotharakos} showed that for the highest-$\dot{E}$ pulsars, $E_c$ begins to fall below the maximal radiation-reaction-limited value, signaling the onset of pair-regulated screening once the equatorial current sheet exceeds a critical compactness threshold. This enhanced in-situ pair creation softens the acceleration, leading to a saturation of $E_c$ near a few GeV \citep{2025kalapotharakos} and to broader, flatter SED peaks \citep[][]{Smith23}. To capture this behavior, we impose a saturation at $E_c=2.4\,\mathrm{GeV}$, producing a second knee around $\dot{E}\sim 10^{34}\,\mathrm{erg\,s^{-1}}$, and include a $0.1$\,dex Gaussian scatter to reflect the observed dispersion.
The resulting realization of $E_c$, for one synthesized population, is
shown in the left panel of Figure \ref{fig:fp}.

To further improve the fidelity of the MSP spectra, and potentially to better extrapolate below the LAT band, we also allow
evolution of the spectral shape
with $\dot{E}$.  This evolution was explored empirically in \citetalias{Smith23} in the
context of the PLEC4 model, where the $\Gamma$ and $d$ parameters
correspond to the first and second logarithmic derivatives at the
scale energy, $E_0$, and where larger (smaller) values of the $b$ parameter further sharpen (broaden) the spectral peak; see Figure 15 and text of \citetalias{Smith23} for a detailed
presentation.  Without loss of generality, we
set $E_0=E_c=E_p$, where $E_p$ is the peak energy, so that $\Gamma$
and $d$ are the derivatives at the spectral peak, $\Gamma_p$ and
$d_p$.  We adopt two \citetalias{Smith23} results:
$\log_{10}b=0.42-0.15\log_{10}\dot{E}_{36}$, and $d_p = 0.69
-0.16\log_{10}\dot{E}_{36}$, For low $\dot{E}$, we saturate these 
relations at $b=1$ and $d=4/3$, which is a monenergetic curvature
spectrum.
This prescription fully determines the PLEC4 model shape, which
we adopt for our simulation.

To determine the normalization, we estimate $\mathcal{L}_{\gamma}$ using
the theoretically-expected relation reported in
\citet{Kalapotharakos22}, specifically
$\mathcal{L}_{\gamma}= 2.4\times10^{33}
E_{c,\mathrm{GeV}}^{4/3}\,B_8^{1/6}\,\dot{E}_{36}^{5/12}$\,erg\,s$^{-1}$.
We tuned the normalization
of this expression to yield good agreement between the
synthesized MSPs and the observed $\gamma$-ray MSP population.  In
order to encapsulate beaming effects and other physics (e.g., unknown
magnetic field obliquity and configuration, and varying moment of inertia), we add
0.3\,dex of gaussian random noise to $\mathcal{L}_{\gamma}$.  One realization of
the resulting distribution of pulsar luminosities is shown in the
right panel of Figure \ref{fig:fp}.

\begin{figure}
\centering
  \includegraphics[angle=0,width=\linewidth]{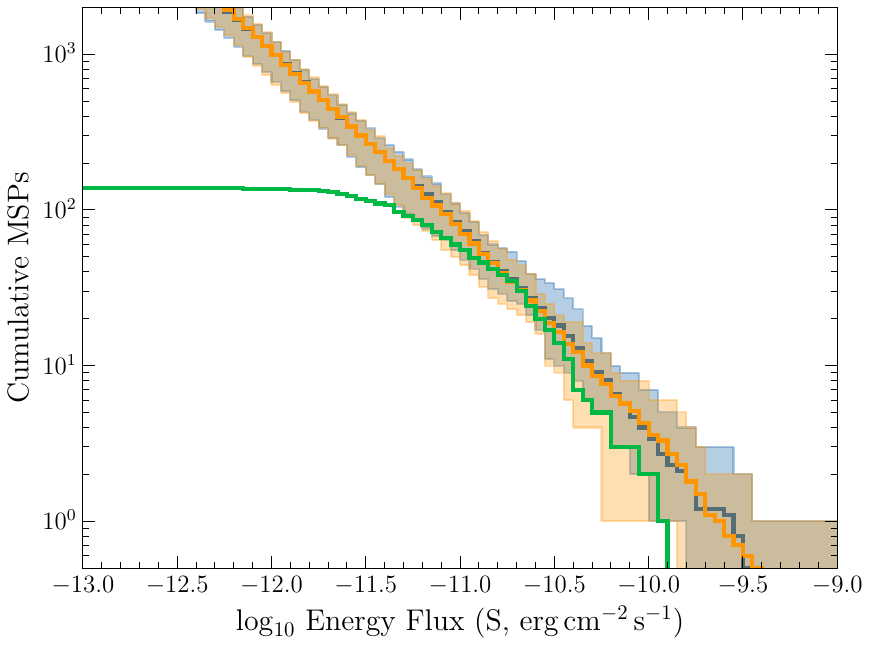}
  \caption{\label{fig:logNlogS}The cumulative distribution of
  $\gamma$-ray fluxes observed at earth as predicted from the
  fundamental plane relation for 10 synthesized MSP populations.  The blue trace
  indicates the mean over the realizations and includes additional scatter in $E_c$ and $\mathcal{L}_{\gamma}$, while
  the orange trace gives the results without this scatter.  The envelope of the same color indicates
  the minimum and maximum values encountered over the 10 realizations.  The green trace is the observed flux
  distribution of LAT MSPs.}
\end{figure}

Since we have encapsulated beaming effects in the $\mathcal{L}_{\gamma}$ scatter, we convert these luminosities to fluxes simply as $F_{\gamma}=\mathcal{L}_{\gamma}/(4\pi\,d^2)$, where $d$ is the
pulsar distance from the population synthesis.  
Figure \ref{fig:logNlogS} shows the resulting distribution of LAT-band energy
fluxes, $S$, both with and without the additional scatter on $E_c$ and
$\mathcal{L}_{\gamma}$.  Such scatter can induce a shallower slope in
the large-$S$ limit (where there are few objects), but it is clearly
not a strong effect in this case.  Although the synthesis slightly overpredicts the number of bright MSPs, given that there is essentially only one free parameter in the model---the normalization---the model is excellent.

%


\subsection{Caveats}

We made several implicit assumptions in calibrating to the observed population, which may not hold in extrapolation to the bulge AIC scenario\footnote{The contribution of AIC to MSPs in the disk is small \citep{2025arXiv251015661S}.}, or below the knee $\dot{E} \lesssim 3\times 10^{32}$~erg~s$^{-1}$. The $\gamma$-ray luminosity of pulsars is collimated, and the solid angle of the $\gamma$-ray beam depends on the magnetic obliquity of the pulsar. Pulsars which are nearly aligned rotators have narrow beams, while highly orthogonal rotators illuminate nearly the full sky \citep[e.g.,][]{2010ApJ...714..810R,2012A&A...545A..42P,2014ApJ...793...97K,2014ApJS..213....6J,2015A&A...575A...3P,2019MNRAS.490.1437V,2021A&A...647A.101B,2021A&A...654A.106P,2023ApJ...954..204K,2024ApJ...962..184C,2025MNRAS.541..806I,2025A&A...695A..93C}. Thus, in a phase-averaged formalism at the population level, a constant beaming factor that converts observed flux to luminosity implicitly encodes the distribution (mean, and scatter) of magnetic obliquities in the population of MSPs and its impact on $E_c$, spectral width and index, and pulsed fraction. The observed distribution of beaming factors (or magnetic obliquities) is also biased against detection of potentially abundant populations with narrow beams (e.g. near-aligned rotators). 

Little is known about the formation and evolutionary channels of magnetic fields in MSPs and the resultant distribution of magnetic obliquities for MSPs in the GC region, particularly for the AIC channel. Thus the scatter of simulated $\gamma$-ray luminosities may be different than assumed. Moreover, if, for instance, AIC has a preference to produce aligned rotators \citep[e.g.,][]{2025arXiv251015661S}, the number of true MSPs will be underestimated from the normalization of the GCE luminosity. This can be assessed with detailed light curve fitting (or even phase resolved spectroscopy and polarization) of a detected GC MSP population, enabled by a future $\gamma$-ray instrument.


Finally, we have made a major but conservative assumption that no additional emission components become apparent at MeV energies, e.g., from secondary resonantly-excited $e^\pm$ pair synchrotron emission \citep{2021ApJ...923..194H}. The form of pulse profile shapes in this MeV regime is also not known, which may impact detectability. However, extrapolation of spectra from NuSTAR observations suggests some energetic MSPs may have pulsed MeV components \citep{2017ApJ...845..159G}, compatible with such pair synchrotron emission (similar to that seen in young energy pulsars such as the Crab). Thus, an observatory with MeV capability may, in practice, detect and time many more MSPs than we forecast from extrapolation of GeV curvature radiation from primaries.

\section{Simulating the Instrumental Response}
\label{sec:virtuallat}

\begin{figure}
\centering
  \includegraphics[angle=0,width=\linewidth]{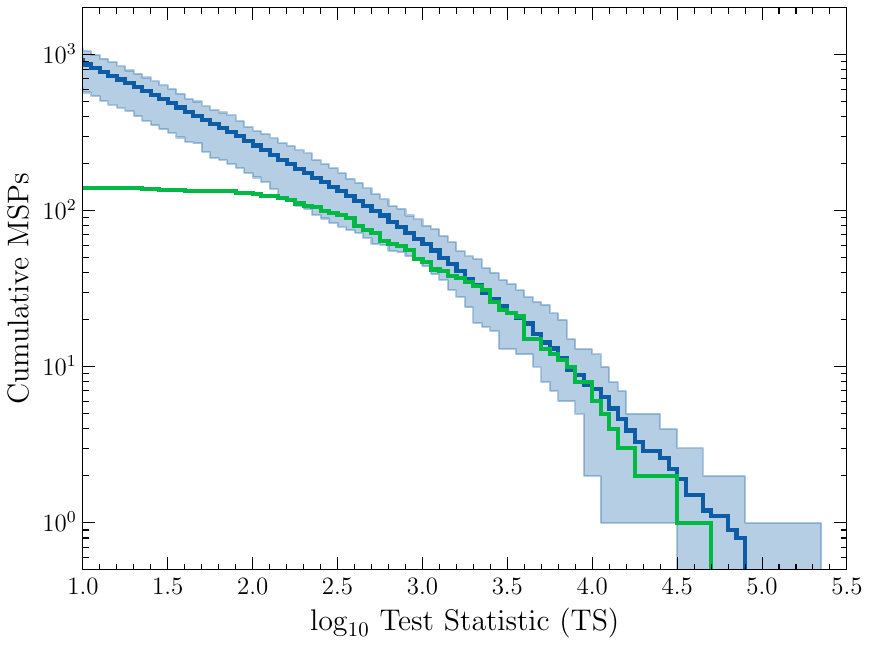}
  \caption{\label{fig:ts_calib}As in Figure \ref{fig:logNlogS}, the
  blue trace and envelope give the mean and range over MSP population
  realizations of the statistical significances (TS) achieved using the virtual LAT described in the main
  text.  The TS values have been scaled to a data set length of 12 years.
  The green trace again gives the
  distribution of observed values. }
\end{figure}

Our goal is to forecast the sensitivity of future $\gamma$-ray
instruments to GW via pulsar timing.  To do this, we
have developed an approach that enables computation of the statistical
significance of pulsar signals using only high-level
specifications of instrument performance.  For example, an instrument
concept might specify an angular resolution over a target
energy range, but it is unlikely to specify the detailed shape of the
point-spread function.

\subsection{Calculating Unpulsed Detection Significances}
\label{sec:unpulsed_ts}

To anchor the technique, we have reduced the detailed LAT instrument response
to such a high-level description and then recorded the
difference in source significances obtained when using
the full instrument response.  This calibration against the
``virtual'' LAT then provides a scaling factor when can be used to predict detailed performance from coarse specifications.

Reconstructed \citep{Atwood13} LAT events can be divided
into four classes, each with its own energy-dependent effective area and detailed PSF model.  This
division allows the photons with the best angular resolution to
constrain source positions.  
The production and characterization of such event
classes reflect a detailed knowledge of the instrument and its
performance, so we re-imagined the LAT ``on the drawing
board'' by combining these event classes with a single type and a simple PSF.  Specifically, characterized the PSF with a single power-law scaling,
$\log_{10}\,(r_{68}/1\,\arcdeg) = -0.01 -
0.715\log_{10}(E/\mathrm{GeV})$, which we determined from a fit over
the energy range of interest ($<$30\,GeV).  $r_{68}$, the radius that
contains 68\% of the photons from a point source, is a common
performance metric.  We likewise summed the effective area for the
four classes.    Finally, we ignore
dependence on the source position within the instrument field-of-view.
This is essentially comparable to choosing an overall exposure
scaling, since the effective area decreases towards the edge of the
field-of-view, and the total exposure for a particular source will
depend on the pointing or survey strategy.  This exposure scaling
factor, which converts source intensities into counts, is the primary
means of calibration of the virtual instrument.  Given a simulated pulsar position and spectrum, we can use this
exposure to calculate the differential expected counts.

We also adopt a simple background emission model.  We consider only
astrophysical backgrounds, as instrumental backgrounds (e.g.~from
activation and cosmic-ray interaction, see \citet{Cumani19}) and local backgrounds (e.g.~from the earth limb/albedo) depend strongly on specific instrument
design\footnote{Moreover, a successful design typically minimizes
instrumental backgrounds compared to astrophysical ones.}.  For pair
regime concepts, the primary background is diffuse Galactic emission,
which we realize with the 4FGL Galactic interstellar emission model \citep{Acero16},
specifically \texttt{gll\_iem\_uw1216\_v13.fits} \footnote{\url{https://fermi.gsfc.nasa.gov/ssc/data/access/lat/14yr_catalog/}}.  For
the low-energy extension (to roughly 10\,MeV) we consider, we
extrapolate the model from its minimum energy (50\,MeV) to $<$1\,MeV.  We verified that the
low-energy turnover is similar to GALPROP models \citep{Porter22} that
extend to 10\,MeV, so this is not likely a major source of
uncertainty.

We also include an isotropic component, e.g. from unresolved emission
from distant blazars, using a
simple power law $dN/dE=0.95\times10^{-7}(E/100\,\mathrm{MeV})^{-2.32}$\,ph\,cm$^{-2}$\,s$^{-1}$\,MeV$^{-1}$\,sr$^{-1}$
\citep{Ackermann15}.  This component is substantially brighter than the Galactic diffuse emission in the Compton regime.

With predictions for the differential source and background counts, we
can estimate the source statistical significance via its
$\mathrm{TS}\equiv2\times\delta\ln\mathcal{L}$, viz.~twice the
difference in the log likelihood obtained with and without a model
containing the source.  Rather than evaluating this quantity
explicitly, in keeping with our simple approximations, we
approximate the TS by evaluating the source and background counts
within $r_{68}$ and then integrating the quantity $S(E)^2/[S(E)+B(E)]\approx\sigma^2(E)$,
over the specified energy range to yield, approximately, the total TS.  Differences
between the likelihood approach and this approximation are further
minimized by the calibration of the virtual instrument to LAT TS value demonstrated below.

Figure \ref{fig:ts_calib} shows the distribution of
the statistical significance calculated in this way for 10 synthesized
MSP populations (each receiving a weight of 1/10).  There is excellent
agreement with the values of actual LAT pulsars, which have been
calculated using the full response.  At a TS threshold of 100, the
average number of MSPs detected is 261 (see Table
\ref{tab:gev_concepts}), which is in agreement with the LAT
population\footnote{The number of currently detected MSPs, about
half of this value, is certainly incomplete.  See \citet{Shawaiz25}
for a recent discussion of the selection efficiency.}.
  
At a high level, a target instrument might differ in effective area, field-of-view, energy range, and/or angular resolution.  For MSPs, which are distributed widely over the sky, it is the product of effective area, $A$ and field of view, $\Omega$---the acceptance, or grasp---that determines the performance for MSP discovery and pulsar timing.   Thus, for many of the concepts we consider below, we simply apply the technique outlined above to a target instrument by scaling the LAT effective area, assuming the same field-of-view, and specifying the appropriate $r_{68}(E)$, and we denote the acceptance scaling factor $A\Omega$.  The energy range is not so easily treated, but we 
foreshadow some of the following results and note that the most
important energy range of $\gamma$-ray PTA science is roughly
0.1--10\,GeV, and this means the $A\Omega$ scaling of LAT captures the most scientifically interesting cases.


\subsection{Converting Unpulsed Significances to Timing Precision}
\label{sec:timing}

The development so far enables the accurate simulation of a population
of unpulsed MSP significances for a high-level instrument concept.
The precision with which a pulsar can be timed depends on its overall
signal-to-noise ratio (S/N), the pulse profile shape (sharp profiles
concentrate photons into a smaller time window, increasing S/N), and
on the spin frequency.  Here, we show that these properties can be
directly related to the unpulsed significance.

The pulsed significance of a particular pulsar will reflect the shape
of its pulse profile.  It can be evaluated in multiple ways, e.g.
using the H-test \citep{deJager89,Kerr11} or with a template method
using an analytic model
for the pulse shape as a function of pulse phase $\phi$, $f(\phi)$, in
which case the log likelihood (compared to no pulsations, $f(\phi)=1$)
is $\ln\mathcal{L}=\sum_i \ln \left[w_i f(\phi_i) + (1-w_i)\right]$,
where $w$ is a photon weight \citep{Bickel08,Kerr11} giving the
probability a photon originates from the pulsar or the background.

For the observed LAT population, the ratio of the pulsed to unpulsed significance is about 1, no matter which pulsation statistic is used.  The typical ratio for young pulsars (MSPs) is slightly greater than (less than) unity, which is likely to due the wider effective widths of MSP pulse profiles.  This means that, on average, the shapes of $\gamma$-ray pulse profiles do not strongly influence the performance of a pulsar as a clock.  Practically, this means that we can determine the timing precision by scaling the unpulsed significances appropriately.

\begin{figure}
\centering
  \includegraphics[angle=0,width=0.98\linewidth]{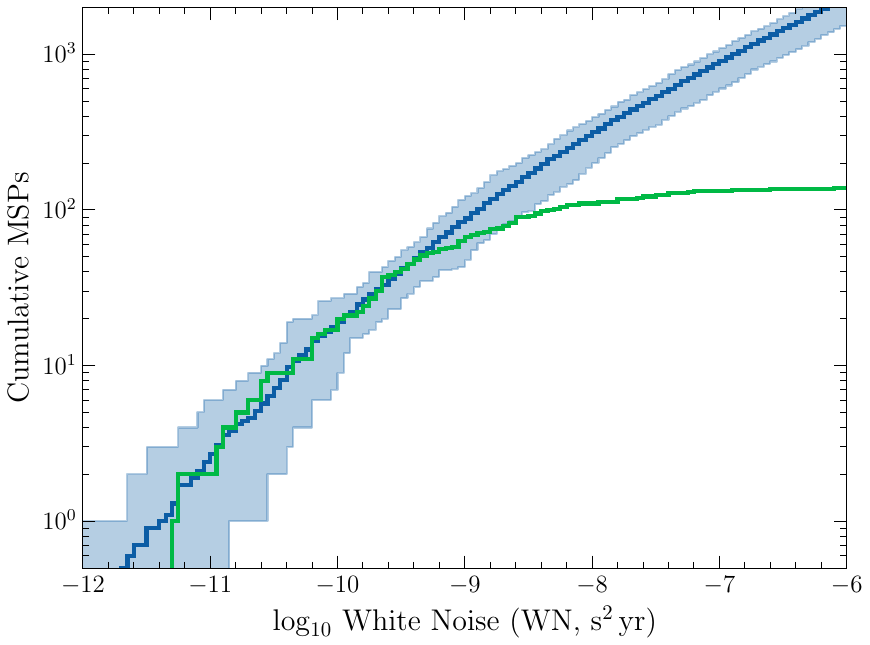}
\caption{\label{fig:wn_calib}The distribution of white noise (WN)
(essentially pulsar timing precision) for 10 realizations of the disk
  MSP population in the virtual LAT.  As in previous plots, the solid line indicates the mean while the shaded region indicates the range of the simulations.  The green trace gives the distribution of observed WN values.}
\end{figure}

We do this scaling by predicting the white noise (WN) amplitude for MSPs.  WN characterizes the performance of a pulsar as an ideal clock, is proportional both to the spin frequency and the source significance, and is better for sharp pulse profiles.
Expressed in units of s$^2$\,yr, WN corresponds to the maximum amplitudes of unmodelled sinusoidal
modulations that could be present in the data.  More concretely, if
the data were synthesized into idealized pulse times of arrival
(TOAs) collected with a cadence of 2\,yr$^{-1}$, then $\sqrt{\mathrm{WN}}$
corresponds to the TOA precision.
The WN in the LAT sample is related to the unpulsed significance by both a scale factor and a scatter, the latter effectively marginalizing over distributions in spin frequency and pulse profile shape.  We measured the scale factor from the data, and we determined that 0.6\,dex of scatter produced good agreement between the synthesized population and the LAT sample.  We show a realization of WN from a synthesized MSP population in Figure \ref{fig:wn_calib}.

This completes the mapping from synthesized MSP population to GPTA measurement set.

\begin{figure*}
\centering
  \includegraphics[angle=0,width=0.49\linewidth]{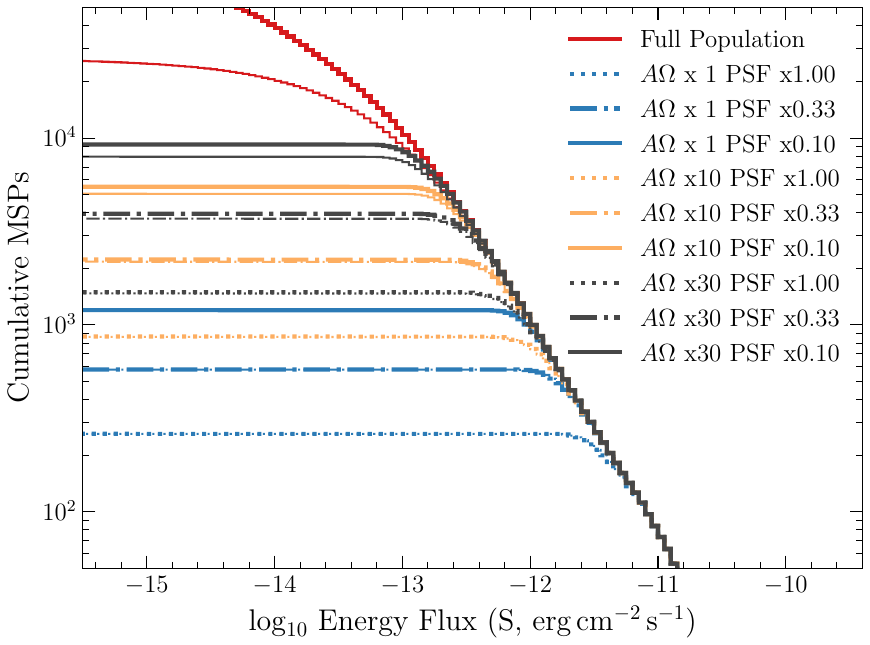}
  \includegraphics[angle=0,width=0.49\linewidth]{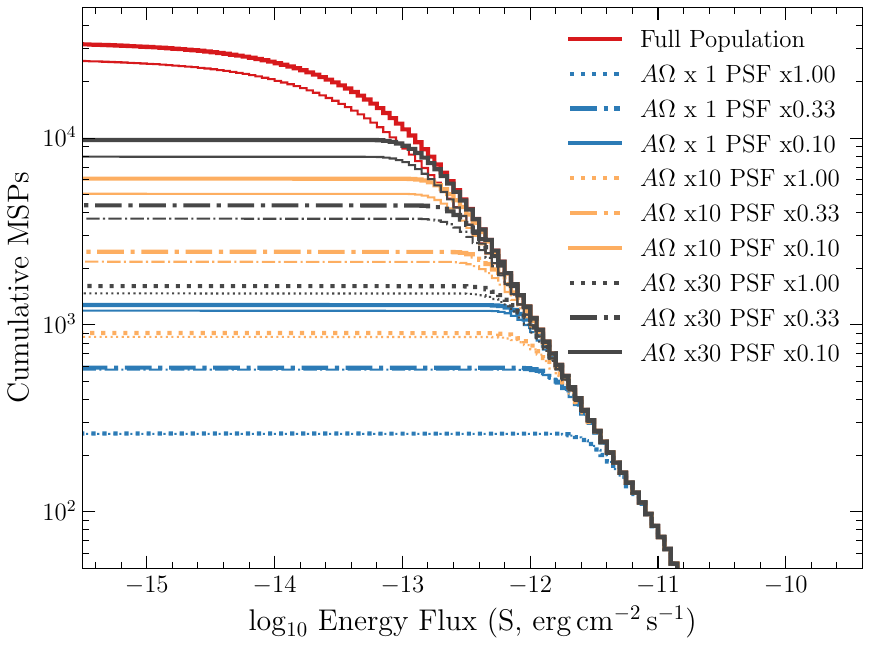}
  \caption{\label{fig:gev_logNlogS}The distribution in energy flux for
  detected (TS$>$100) MSPs in the disk+bulge population scenarios S2 and S3, and for the nine GeV-band concepts
  discussed in the text (heavier lines).  Line colors and styles distinguish the 9 concepts.  The disk-only contributions, i.e. S1, are also shown (lighter lines).   Left: In S2, the bulge
  population comprises $1.2\times10^5$ MSPs from
  an accretion-induced collapse pathway, and it is evident that relatively few of these bulge pulsars are detectable.
Right:
  In S3, the bulge
  population comprises 6,700 MSPs assumed to share the same
  $\gamma$-ray emission properties as the disk.  Due to their higher
  individual $\gamma$-ray luminosity, these pulsars are more detectable, particularly by instruments with a good PSF.}
\end{figure*}

\section{Forecasting PTA Performance}
\label{sec:ptaperf}
To assess the capability of the resulting pulsar timing data sets to constrain the
amplitudes of GW sources, we use the package
\texttt{hasasia} \citep{Hazboun2019Hasasia} as follows.  For a
synthesized population, we calculate the $\gamma$-ray spectra and
expected TS for the virtual instrument in question following the
procedure outlined above.  We select pulsars with $\mathrm{TS} > 100$
and $\mathrm{WN} < 10^{-8}$\,s$^2$\,yr and
convert the estimated WN into a set of virtual TOAs with cadence
12\,yr$^{-1}$ with a uniform uncertainty of $\sqrt{\mathrm{WN}/6}$.  Finally, we
ingest those pulsars, along with the synthesized sky positions, into
\texttt{hasasia}, which estimates a strain sensitivity spectrum
for each pulsar and, with the appropriate overlap functions, the
sensitivity to an incoherent background or a specific GW source.

In principle, sensitivity to the GWB is limited both by spin noise in
MSPs, and by the GWB itself (self noise).  However, the distribution
of intrinsic spin noise over the MSP population is not yet well known, and in at least some pulsars the noise may
be dominated by the GWB itself \citep{Miles25}.  Consequently, the
sensitivity calculations we provide account only for white noise.
They can be adjusted post-facto to account for spin-noise and GWB
self-noise, while the converse is not true.  Furthermore, for direct
comparison with the results of \citet{Ajello22}, we assume a
data set duration of T$_{\mathrm{obs}}=12$\,yr.  The results can be
re-scaled to different data spans as $\mathcal{A}_{\mathrm{gwb}}\propto
T_{\mathrm{obs}}^{-13/6}$.

\section{Pair Production Telescopes}
\label{sec:gev}

At MeV energies, photons interact with matter primarily through
Compton scattering, where the scattered photon and Compton electron
must be detected and characterized to reconstruct the $\gamma$-ray
properties.  The conversion of photons into $e^{\pm}$ pairs becomes the dominant interaction process between a few and about 30\,MeV, and tracking of
the charged products is relatively easier.

\subsection{LAT-like Instruments}

The main concepts we consider are a series of LAT-like instruments, and we briefly motivate this choice.  First, known $\gamma$-ray MSPs emit most of their power in the 0.3--3\,GeV decade.  An analysis below confirms that a detailed understanding of future instrument performance outside of this range is unimportant, so it suffices to scale to LAT.  Second, acceptance\footnote{Recall that for the purposes of PTAs,
effective area and field-of-view are interchangeable.} is critical for monitoring a large MSP population, so scaling to the LAT field of view also makes sense.  Finally, as we argue below, the physics of particle tracking largely fixes properties of the PSF, so that the energy dependence of the LAT PSF is nigh-universal.  The PSF is particularly important, because for
background-limited source detection, the threshold energy flux
$S_{\mathrm{min}}^{-1} \propto N_{\mathrm{src}} /
\sqrt{N_{\mathrm{bkg}}} \propto
\sqrt{A_{\mathrm{eff}}\,T_{\mathrm{obs}}}/r_{68}$.  Because weight and
cost often scale as $A_{\mathrm{eff}}$, an improved PSF is a promising
avenue for detecting the more distant disk MSPs and for clarifying the
origin of the Fermi GCE by constraining the bulge MSP population.

In a tracking detector, the $e^\pm$ position will be measured many times along its track, 
allowing an optimal reconstruction using a Kalman filter~\citep{1987NIMPA.262..444F}, 
which has angular resolution $\sigma_\theta \simeq \sigma_0 E^{-3/4}$~\citep{2023arXiv230519690B}. 
This $-3/4$ value is very close to that measured for the LAT, $-0.715$.
So the energy scaling should be similar for every implementation of this concept, 
the differences in the choice of conversion material and detection technique 
affecting only the $\sigma_0$ constant. 
Other concepts could drastically alter certain
properties.  E.g., the addition of a coded mask could provide substantially
improved and effectively energy-independent angular resolution for
$E<1$\,GeV.  However, the energy-scaling of the LAT-like PSF compared
to a flat PSF over a single decade will not drastically change the PTA
performance, so the performance of such concepts can be roughly derived from results tabulated here.



The concepts considered are summarized in Table
\ref{tab:gev_concepts} and are permutations of instruments with
1$\times$, 10$\times$ and 30$\times$ the acceptance ($A\Omega$, see \S\ref{sec:unpulsed_ts}) of LAT; and with a PSF
that is identical, or for which $r_{68}$ has the same energy
dependence but is 3$\times$ smaller and 10$\times$ smaller.




\begin{deluxetable}{c||c|rrr|c}
  \tablehead{
    \colhead{Index} &
    \colhead{Configuration} &
    \colhead{$N_{\mathrm{disk}}$} &
    \colhead{$N_{\mathrm{bulge}}$} &
    \colhead{$N_{\mathrm{bulge}}$} & 
    \colhead{$A_{15}$} \\
    \colhead{} &
    \colhead{$A\Omega\times$, PSF$\times$} &
    \colhead{S1} &
    \colhead{S2: AIC} &
    \colhead{S3: Disk} &
    \colhead{}}
\startdata
  1& \hphantom{0}1, 1.00 &   261 &    0 &    1 & 10.5 \\
  2& \hphantom{0}1, 0.33 &   577 &    1 &   14 &  7.4 \\
  3& \hphantom{0}1, 0.10 &  1193 &    8 &  86 &  5.7 \\ \hline
  4&            10, 1.00 &   863 &    4 &   43 &  3.0 \\
  5&            10, 0.33 &  2181 &   60 &  285 &  2.1 \\
  6&            10, 0.10 &  5037 &  456 & 1035 &  1.6 \\ \hline
  7&            30, 1.00 &  1475 &   23 &  140 &  1.6 \\
  8&            30, 0.33 &  3707 &  234 &  661 &  1.1 \\
  9&            30, 0.10 &  7953 & 1297 & 1827 &  0.9 \\
\enddata
  \tablecomments{\label{tab:gev_concepts} The concepts for future GeV
  instruments along with expected results from the three population
  scenarios.  The first row corresponds to the virtual LAT, on which
  the scaling is based,
  and the predicted GWB sensitivity and number of detected
  MSPs is in good agreement with the observational results.  The $N$
  columns give the mean number of
  pulsars detected above a threshold of $\mathrm{TS}=100$ for the disk
  and for the two bulge-MSP scenarios.  Because the bulge MSPs are
  much fainter, the GWB sensitivity depends primarily on the disk, and
	so we conservatively report results   
  only for a PTA composed of disk MSPs (S1).
  The scatter on $N$ over the simulation realizations is about
  15--20\%, though clearly larger when $N\lesssim10$.  The scatter
  on $A_{\mathrm{gwb}}$ is about 25\%.}
\end{deluxetable}

\begin{figure}
\centering
  \includegraphics[angle=0,width=0.98\linewidth]{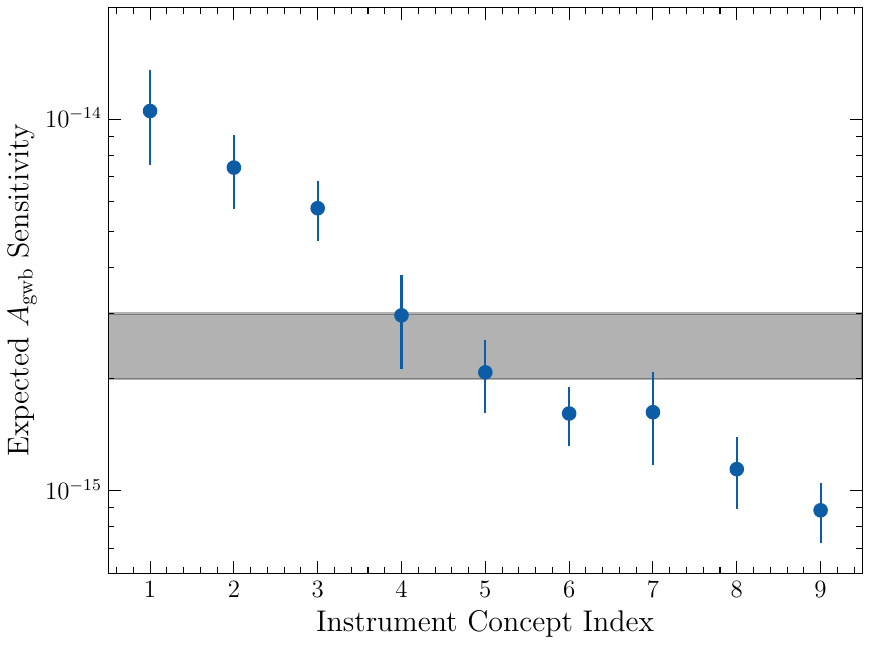}
  \caption{\label{fig:gev_gwb}The limiting $A_{\mathrm{gwb}}$ (at $f=1$\,yr$^{-1}$) that
  could be detected with each of the nine $\gamma$-ray concepts discussed in the
  main text and Table \ref{tab:gev_concepts}.  Only the results from
  scenario S1 are shown here, since S2 and S3 differ very little.  The
  gray shaded region gives the approximate range of candidate GWB
  signals currently detected by radio PTAs,
  $\mathcal{A}_{\mathrm{gwb}}=2$--3$\times10^{-15}$. Performance below the gray band enters the self-noise regime from the GWB, and other features of the PTA (e.g. sky coverage, number of pulsars) become consequential.}
\end{figure}

To assess these concepts, we determined the mean number of MSPs detected over the 10 realizations of S1, S2, and S3 (see \S\ref{sec:bulge} for details of
these populations) and report them in Table~\ref{tab:gev_concepts} and Figure~\ref{fig:gev_logNlogS}.  We further estimate the performance for GW detection, again listed in Table
\ref{tab:gev_concepts} and depicted in
Figure \ref{fig:gev_gwb}.

\subsection{Towards the Pair Production Threshold}

The natural low-energy cutoff of the pair regime is the energy at which 
Compton scattering becomes the dominant interaction process, which is roughly
130\,MeV in hydrogen, 28\,MeV in carbon, 15\,MeV in silicon, 
and 5\,MeV in tungsten.
One clear way in which the LAT could be improved is in its
low-energy performance, which is limited by the requirement of a signal in three consecutive tracker planes to trigger a read out of the instrument.
However, a detector in which conversion happens in a less dense material, 
with improved electronics and relatively more active material, could detect 
 $<$100\,MeV events much
more efficiently \citep[see e.g. the predicted effective area of
VLAST,][]{Pan24}.  We therefore consider a near-ideal pair instrument with an effective area that rapidly increases from 20 to 30\, MeV and that saturates at 100\,MeV to the same efficiency it has at 1\,GeV.  Under
the assumption the PSF remains dominated by multiple scattering, we
simply extrapolate the LAT PSF to this energy range.

We performed simulations using the same LAT scalings delineated in
Table \ref{tab:gev_concepts} with the additional 0.01--0.1\,GeV decade
of energy range.  While virtually no new MSPs were detected by the
additional bandwidth, we found improvements ranging from 5\% (low angular
resolution) to 10\% (high angular resolution) in GWB
sensitivity, indicating modestly higher timing precision. 

\begin{figure}
\centering
  \includegraphics[angle=0,width=0.98\linewidth]{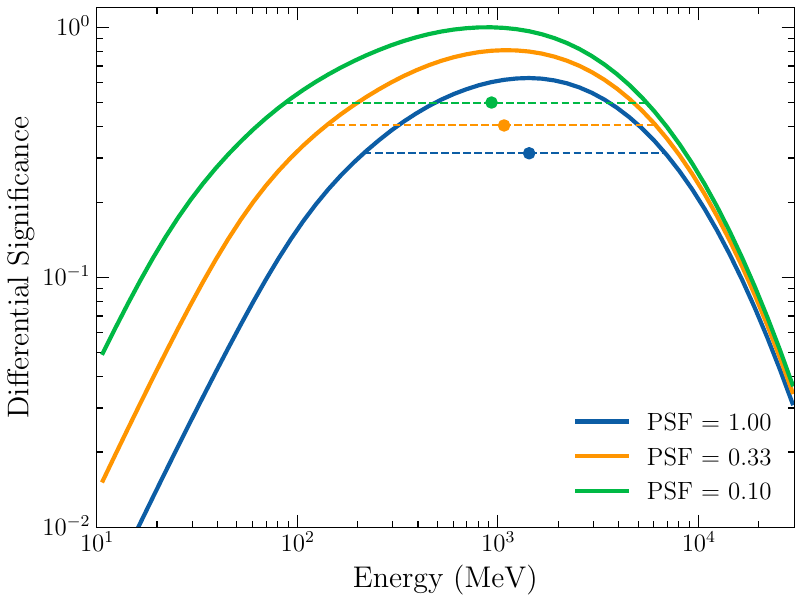}
  \caption{\label{fig:gev_bandpass}The normalized differential source significance (solid lines) for the disk population MSPs, showing the relative TS that would be accumulated at each energy if the instrument effective area were flat.  This identifies the optimal energy range for an instrument with a LAT-like PSF to operate for MSP observations.  The units are arbitrary, but relative differences reflect the increased statistical significance accumulated with increased angular resolution.  The dashed lines indicate the full-width at half-maximum.}
\end{figure}

We can generalize this somewhat by considering the relative rate at which TS is accumulated as a function of energy.  To compute this quantity, we simply assumed a flat effective area and summed the $S^2(E)/(S(E)+B(E))$ for all MSPs in the realized population.  This gives the curves shown in Figure \ref{fig:gev_bandpass}.  These indicate that the ``sweet spot'' begins from 80\,MeV (for the best PSF) to 200\,MeV (for the LAT PSF) and extends to 5 GeV.

While we
thus conclude that the 0.01--0.1\,GeV band is not a high priority for
PTA science, accessing lower-energy spectral measurements would
undoubtedly improve understanding of pulsar emission mechanisms\footnote{For instance, MeV observations could assess the pair synchrotron components as a function of $\dot{E}$, new polar cap emission components, and assess the positron production from pulsars \citep{2019BAAS...51c.379H}.}.

\subsection{Modifying the Gamma-ray Luminosity Function}

Because there are no observational constraints on the $\gamma$-ray
emission of low-$\dot{E}$ pulsars, we also considered a modification
of the $\gamma$-ray emission model, specifically the transition of the
limiting cutoff energy from the radiation-reaction limited regime to
the voltage-limited regime.  If the latter cutoff energies are
increased, e.g. by raising the fraction of the polar cap voltage
tapped $\eta_{PC}$, then the ``knee'' where $E_c$ transitions moves to
lower values of $\dot{E}$ and the efficiency of MSPs with $\dot{E}$
values around $10^{32}$\,erg\,s$^{-1}$ will increase.  We checked the
impact of such a modification by evaluating the scenarios with
$\eta_{PC}=0.65$ (instead of 0.4), which moves the knee down by about a
factor of 3, to $10^{32}$\,erg\,s$^{-1}$.  However, the change in logN-logS
only begins to become apparent around
$S=10^{-14}$\,erg\,cm$^{-2}$\,s$^{-1}$, which is below the threshold
of any of the GeV concepts presented here (see Figure
\ref{fig:gev_logNlogS}).  Consequently there is
little sensitivity of the results presented here to the specific
formulation of the $\gamma$-ray emission mechanism below the range
that is calibrated by the observed $\gamma$-ray MSP population.

\section{Compton Instruments}
\label{sec:mev}

We also consider instruments operating in the Compton regime, which
extends from roughly 0.1--10\,MeV.  This spectral window is
notoriously difficult and famously underexplored, earning it the
moniker ``The MeV Gap''.  
The MeV spectra of MSPs are essentially
unconstrained, nor do we have a contemporary instrument like
the LAT to use to calibrate future instrument concepts.  (For comparison with the LAT, the NASA Small Explorer mission 
COSI~\citep{Tomsick19}, which will launch in 2027, has been optimized for energy resolution and has an effective area of about 10\,cm$^2$.)  Thus, we
begin this section with the strong caveat that in the absence of
observational constraints the results depend entirely on our
assumptions.

For the MeV spectra, we simply extrapolate the PLEC4 models from the
pair regime.  We tested two different extrapolation methods: one in
which we used the \citetalias{Smith23} scaling relations for $b$ and $d_p$ described
above, and a second in which we fixed the asymptotic low-energy
spectral index $\Gamma=2/3$, as for a monoenergetic spectrum.
However, we did not consider any additional emission components.

For the instrument, we surveyed the literature and established a
baseline concept representative of the most sensitive concepts; we
discuss this more in \S\ref{sec:discussion}.  It is a Compton imager
with an effective area of 15,000\,cm$^{2}$ operating from 0.3--3\,MeV
with an angular resolution of r$_{68}$=2\arcdeg. Because the
isotropic background is so intense, we also considered adding a coded
mask in order to achieve an angular resolution of 0.1\arcdeg.

Unfortunately, subject to these assumptions, neither concept is well suited
for MSP pulsar timing.  The baseline concept essentially detects no
MSPs, while the addition of the coded mask allows detection of 20--40
MSPs, depending on the assumptions made about the PLEC4 model.

\begin{figure*}
\centering
  \includegraphics[angle=0,width=0.49\textwidth]{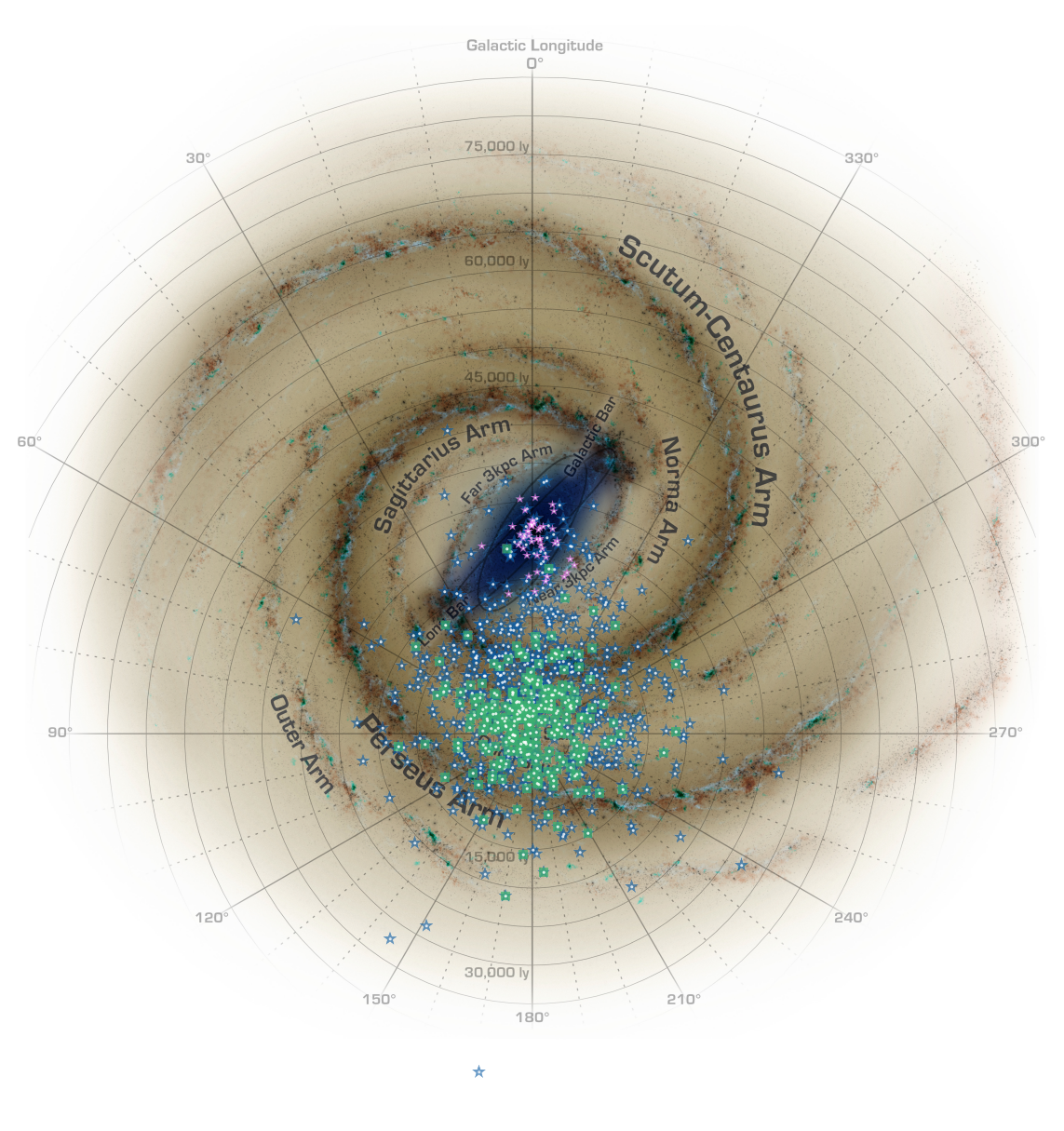}  \includegraphics[angle=0,width=0.49\textwidth]{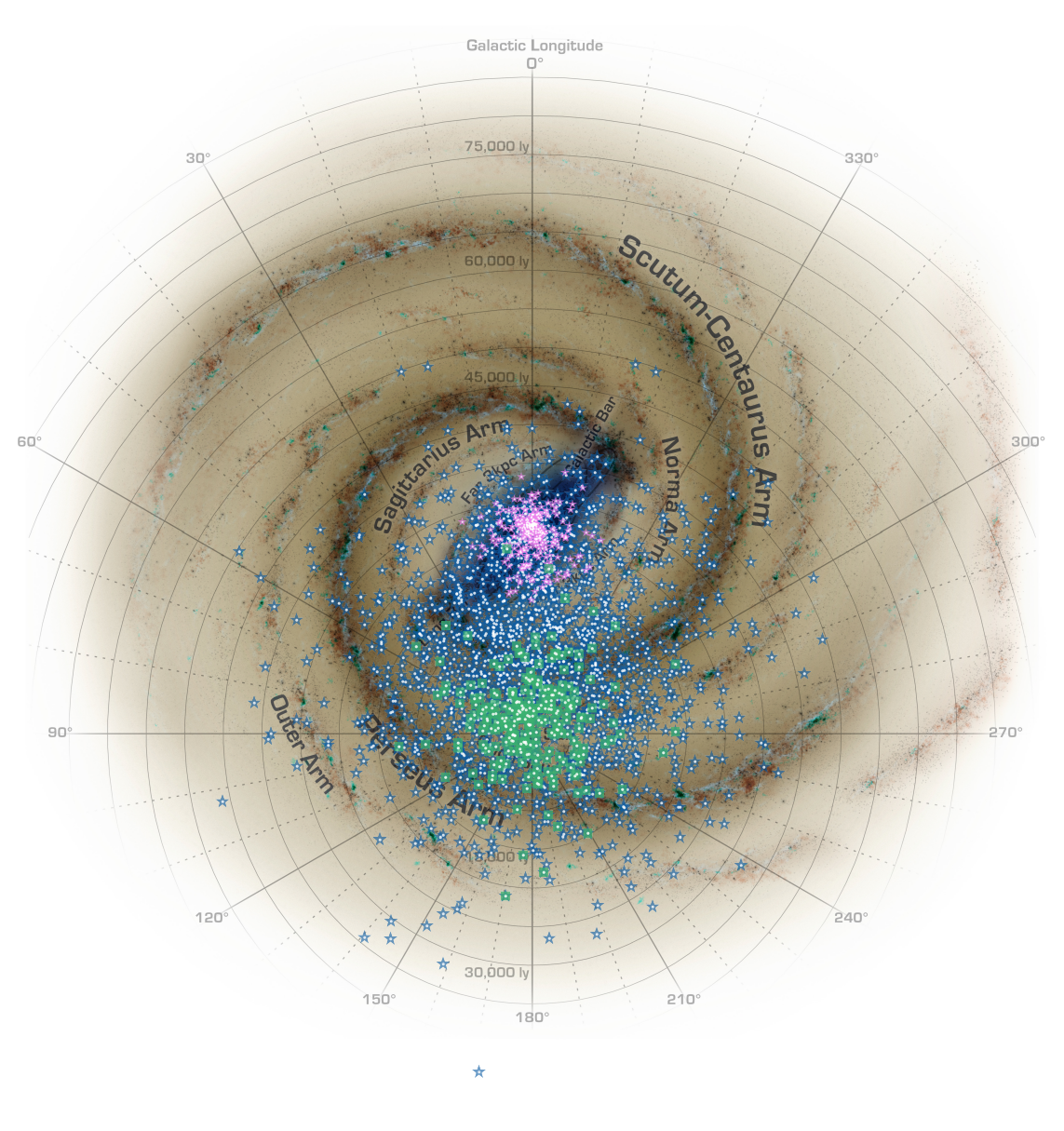}
  \includegraphics[angle=0,width=0.49\textwidth]{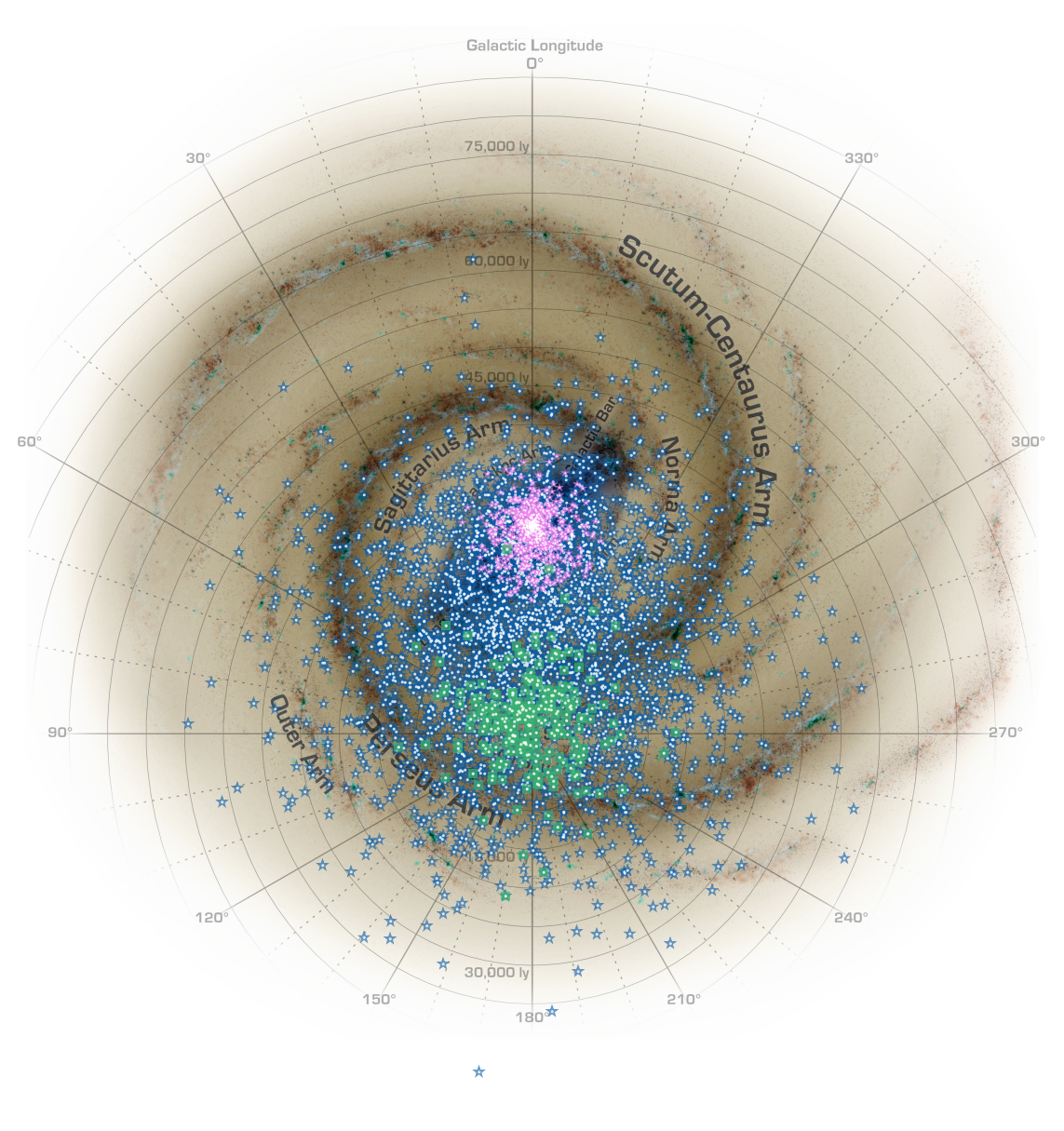}
  \caption{\label{fig:topdown}The positions of detected, simulated MSPs in the Galaxy.  Each panel follows the same scheme and gives MSPs from concepts 4, 5, and 8.  The green points are those detectable by Concept 1 and are all disk MSPs.  The blue points indicate detected disk MSPs, and the lilac points bulge MSPs under S3.  Comparing Concepts 4 and 5, the impact of an improved PSF is apparent.  Concept 8 improves on Concept 5 by increasing the acceptance by a further factor of 3.}
\end{figure*}

\section{Discussion}
\label{sec:discussion}

We have developed a successful model of MSP pulsar emission, validated
it against LAT pulsars, and used it to predict the MSP discovery rates
and pulsar timing performance of a set of baseline configurations for
future $\gamma$-ray instruments.  We now discuss the implications of these results and the potential impact of future GPTA observations.


\subsection{Synergy with Radio PTAs}

We have focused on the capabilities of a $\gamma$-ray PTA operating independently.  However, just as radio PTAs achieve their best results by combining the data sets of individual PTAs, so too will the GPTA enhance and be enhanced by radio PTAs.  One important method is in providing independent estimations of spin noise and GWB models, uncontaminated by the IISM, but this cannot be quantified in the absence of real data.  Here, we discuss contributions from finding new MSPs for radio PTAs to monitor, and from unique information about GW signals.

\subsubsection{Complementarity}
Generally, the best-timed $\gamma$-ray pulsars are not the best-timed radio pulsars.  This is because radio fluxes and $\gamma$-ray fluxes are essentially uncorrelated (\citetalias{Smith23}).  One of the best-timed $\gamma$-ray MSPs, J1231$-$1411 is so faint \citet{Cromartie25} that timing it with the sensitive MeerKAT instrument requires a very large investment of observing time.
And with only a few exceptions \citep[e.g.][]{Guillemot12}, the pulse profiles of radio and $\gamma$-ray pulsars are different.  Thus, because there are a finite number of MSPs in the Galaxy, this means $\gamma$-ray pulsar timing can provide constraints on GW sources that are otherwise inaccessible, no matter how sensitive radio telescopes become.

\subsubsection{An MSP Treasure Map}

The current number of known MSPs outside of globular clusters is
roughly 600, which includes substantial recent contributions from FAST \citep[e.g.][]{Han25}.  Future surveys with powerful radio telescopes will
certainly expand this number.  \citet{Keane15} predicted the discovery of about 2000 MSPs with SKA Phase 1, although these numbers should be revised downward following rebaselining of the SKA. Likewise, \citet{2024AAS...24326104S} predicts about 3000 MSP detections for DSA-2000.

However, radio surveys will inevitably miss many MSPs, for
a variety of reasons.  Like all pulsars, MSPs scintillate in the
ionized interstellar medium, and so at some epochs an
otherwise-detectable MSP will be fainter than the survey threshold.
Similar arguments apply to MSPs in binaries: they are overall more
difficult to detect because pulsar searches must add parameters for
acceleration and/or jerk \citep{Andersen18}, and some epochs are less
favorable due to the particular orbital phase during the observation.
If the wind of a binary MSP is ablating its companion, the stripped,
ionized material can eclipse radio emission, also at particular
orbital phases.  Finally, radio signals are dispersed and scattered,
which increasingly blurs the received pulses for more distant MSPs,
especially those embedded in the Galactic plane.

Because of these biases, it has proven extremely effective to target
pulsar-like LAT point sources for radio followup \citep{Ray12}, and
more than 100 MSPs have been found in this way.  It is inevitable that
a future $\gamma$-ray instrument would spark similar success.  Once a
pulsar is discovered in a targeted radio search, it can be timed (also
in radio) until a timing solution is available.  A timing solution
allows phasing up of the $\gamma$-ray photons and subsequent detection
of the $\gamma$-ray MSP.  The MSP can at this point be added to both
radio and $\gamma$-ray PTAs, with the additional advantage that even
pre-discovery $\gamma$-ray data can be used, since every photon
timestamp is recorded with high precision.  This is a powerful synergy
between radio and $\gamma$-ray observations!

Currently, this iterative approach is the only way, essentially, to
discover binary MSPs using $\gamma$ rays, since it is prohibitively
computationally expensive to carry out a blind search over orbital
parameters \citep{Nieder20}.  However, the combination of new
techniques \citep{Gazith25} and reduced background\footnote{The
lattice technique of \citet{Gazith25} require good separation of
source and background on a per-photon basis, and so greatly benefits
from a better PSF.} could alleviate this problem and allow the direct
detection of many more $\gamma$-ray pulsars.  This happy outcome would
reduce the need for targeted radio followup and allow new MSPs to be more rapidly ingested
into radio PTAs.

\subsection{Detecting and Characterizing Bulge MSPs}

As we argued above, the detection of bulge MSPs does not substantially enhance the performance of the GPTA---they are simply too faint.  However, determining the nature of the MSP bulge population offers substantial insight into the evolution of the Milky Way and potentially a solution to the nature of Fermi GCE.  A detailed analysis is beyond the scope of this paper, but it is worth discussing in broad strokes the discriminating power of these concepts, and in particular highlighting the role of angular resolution.

In order to confirm the presence of a bulge MSP population, a careful analysis of the spatial and spectral distribution of the full MSP population is required.  However, due to the low projected density of the disk population within $\sim$1\arcdeg{} of the Galactic center, we argue that a reasonable threshold for confirming the presence of a bulge population is detection of a few tens of its MSPs.  Most of the concepts (3, marginally 4, 5--9) we presented are capable of this, though in S1 only concepts 5--9 are capable.  In particular, concept 3 outperforms concept 4, trading off a 10$\times$ increase in acceptance for a 10-fold improved PSF.  This trend holds more generally, e.g. comparing concept 5 to concept 7, where a 3-fold improvement in acceptance yields fewer bulge MSP detections than a 3-fold improvement in angular resolution.  We highlight these results visually in Figure \ref{fig:topdown}.

Although we have analyzed only two MSP formation channels, it is clear that most of the concepts are also capable of discriminating between them, yielding insight into the evolution of the Milky Way stellar population.  E.g., concepts 3 and 4 could detect a bulge population in S2 but not in S1.  For all other concepts, the predicted number of detections are readily distinguishable.  Moreover, continued pulsar timing would yield characteristic ages, spindown luminosities, and binary properties, which would much more strongly constrain the formation channels.  Obtaining such observational constraints using radio observations may be much more challenging due to strong scattering in the ISM, and radio detection do not directly yield the $\gamma$-ray luminosity.

\subsection{On Implementation}


We showed that future $\gamma$-ray instruments can function as
powerful pulsar discovery and pulsar timing machines, rivaling
terrestrial radio PTAs.  However, the required performance is a
substantial leap beyond the LAT, which is an extremely successful
NASA Probe-class instrument.  Here, we give an overview of some instrument
designs appearing in the literature and show how they relate to our
baseline concepts, and we also present inchoate but reassuring
sanity checks indicating that even the most performant of the baseline
concepts can be achieved without changing the laws of physics. Note that none of the concepts highlighted below have been specifically optimized for $\gamma$-ray PTA science, and thus a dedicated implementation may substantially outperform. 

The VLAST instrument concept \citep{Pan24} is another
converter/tracker design, though unlike LAT it uses active CsI
converters to enable operation in the Compton regime, in addition to a better energy resolution.  It
achieves roughly 4$\times$ the acceptance of LAT, primarily through
an increase in geometric area of a similar factor.  APT \citep{APT_perf}, also based on long wavelength-shifting fibers, has both increased geometric area and field-of-view relative to LAT, so that operation at L2 (without earth occultation) would allow an increase in acceptance of about 10$\times$.  With the current level of simulation fidelity, both instruments have slightly poorer angular resolution than LAT, but they roughly map to concept 4.

Improved angular resolution can be obtained in the pair regime by
reducing multiple scattering.  Due to the stringent requirements of Compton event reconstruction, observatories focused on the MeV band often provide good angular resolution up to a few GeV, e.g. newASTROGAM \citep{2025arXiv250708133B} and AMEGO-X \citep{Caputo22,fleischhack2021amegoxmevgammarayastronomy}.  We explicitly considered scaling up AMEGO-X, in which conversion
happens in 0.5\% radiation length ($X_0$) layers, c.f. 3.2\% $X_0$ for
the front layers of the LAT tracker.  We characterized the performance
of a $4\times4\times4$ array of AMEGO-X-like converter/trackers,
finding an effective area roughly 4.4$\times$ that of LAT and an
angular resolution improved by $\sqrt{0.5/3.2} = 2.5$.   Thus, this
approach provides a path an instrument somewhere between concepts 2
and 5.  (We note that the aspect ratio of this specific configuration
is twice that of the LAT and so it would likely suffer a reduced
field-of-view, but we have not calculated it at this level of detail, nor optimized the configuration for PTA science.)

GammaTPC \citep{Shutt25} is a liquid argon time projection chamber (TPC)
with charge and light readout on the TPC boundaries.  (See also \citet{GRAMS} for GRAMS.) \
We crudely assess the pair regime performance (see also \citet{2019APh...112....1D}) by forming virtual conversion layers
in the liquid Ar.  We find an effective area similar to its geometric
area, about 70,000\,cm$^{2}$, or 10$\times$ that of LAT.  With a squat, lenticular geometry, its field-of-view likely encompasses the upper hemisphere, leading to an increase in acceptance of 20.
Meanwhile, preliminary estimates of the angular resolution (priv. comm., T. Shutt and B. Trbalic) suggest a PSF of about 1\arcdeg{} at 100\,MeV, several times better than that LAT, though the scaling to 1\,GeV is potentially less favorable.
Thus, GammaTPC or a similar instrument could fulfill our concepts 4/5 or even 7/8.

Achieving angular resolution 10$\times$ better than LAT is difficult, but there are several routes possible.  First, the fine-grained tracking of nuclear emulsion detectors \citep[such as GRAINE,][]{GRAINE} allows such fine resolution, but the timestamping technique only allows 0.1\,s resolution, insufficient for pulsar timing purposes.  (Such a detector might be an excellent pulsar discovery machine, however.)

For high time resolution, an active detector plane can be coupled with a coded mask.  The radiation length of Pb is 5.6\,mm, making a 2.2\,cm thick mask 90\% opaque at 100\,MeV. This makes 0.1\arcdeg{} resolution possible with a mask with 2\,mm feature sizes at a standoff of 2\,m.  (See also \citet{Orlando_2022} for another coded-mask $\gamma$-ray concept, GECCO.)  However, such a thick mask would severely restrict the field-of-view relative to LAT, and a very large detector plane would be required to realize, say, concept 3.  

A third approach is a gas TPC.  We examine the practical limit of this technology with a simple thought exercise.  We consider a TPC of 2-bar Ar gas with a cross-sectional area of 28\,m$^2$ and a pair opacity of 20\%.  This volume fits comfortably in a New Glenn fairing (or may be inflatable with a different launch vehicle), and gives an effective area of about 5\,m$^2$, 7$\times$ that of LAT.  Using only pair tracking for reconstruction of direction and energy \citep{2022hxga.book..101B}, its field-of-view is effectively 4$\pi$, for an acceptance 40$\times$ that of LAT.
Existing gas TPC concepts and prototypes have predicted or achieved
roughly 10$\times$ better angular resolution than LAT at $<$0.1\,GeV
energies \citep[e.g.][]{2014APh....59...18H,Gros18,2019NIMPA.936..405B}.  If the reconstruction-based energy is good enough to preserve efficiency up to a few GeV, it would be sufficient for pulsar detection and timing (Figure \ref{fig:gev_bandpass}).  Thus, a gas
TPC might implement\footnote{We leave the instrumentation of such a
large volume as an exercise to the interested reader.} or even exceed our concept 9.

Finally, there are multiple radio PTA collaborations operating
multiple large radio telescopes.  With this as inspiration, we note
that an increase in acceptance of $N$$\times$ can readily be achieved
by launching $N$ $\gamma$-ray instruments (for instance with a distributed satellite constellation of hundreds of (1) liquid Ar TPC cells or (2) single towers of tracker/converters for silicon-based concepts).

\subsection{Conclusions}

For the purposes of $\gamma$-ray pulsar timing and a $\gamma$-ray PTA in a next-generation $\gamma$-ray observatory,
our results are clear: the pair regime is critical, with the exact bounds depending on angular resolution but roughly extending from 0.1--5\,GeV.  A new, sensitive instrument operating in this regime could contribute substantially to low-frequency gravitational wave astronomy, to pulsar astrophysics, and to understanding the Fermi Galactic Center Excess. On the other
hand, the MeV band is thoroughly underexplored, and strong science
cases have been made to motivate a Compton instrument \citep[e.g.][]{2017ExA....44...25D}.
Fortunately, essentially every instrument concept we examined that can
deliver the required performance in the pair regime also functions as
a superb Compton telescope.  Thus, we offer the following takeaway
message: one can have one's cake and eat it too!

\begin{acknowledgments}
We thank Eric Burns, Chiara Mingarelli, Elizabeth Hays, Casey McGrath, Jacob Slutsky, Tom Shutt, Bahrudin Trbalic, Jim Buckley, Daniel D'Orazio, and Michela Negro for helpful discussions. Z.W. also thanks the FIG-SAG co-chairs Chris Fryer, Paolo Coppi, Tiffany Lewis, Michelle Hui, and Milena Crnogor\u{c}evi\'{c} for motivating this work and useful discussions. The material is based upon work supported by NASA under award numbers 80GSFC21M0002 and 80GSFC24M0006.  
H.T.C.~acknowledges support from the U.S. Naval Research Laboratory.  Basic research in pulsar astronomy at NRL is supported by NASA, in particular via Fermi Guest Investigator award NNG22OB35A. T.C is supported by the NANOGrav collaboration via the National Science Foundation (NSF) Physics Frontiers Center award numbers 1430284 and 2020265.
This work has made use of the NASA Astrophysics Data System. 

\end{acknowledgments}

\software{}
\texttt{hasasia} \citep{Hazboun2019Hasasia}

\bibliographystyle{aasjournalv7}
\bibliography{sr,Future_GPTA}

\end{document}